\documentclass[twocolumn, 10pt, journal]{IEEEtran}

\usepackage{multirow}
\usepackage{epsfig}
\usepackage{subfigure}
\usepackage{cite}
\usepackage{amsmath}
\usepackage{amssymb}
\usepackage{footnote}
\usepackage{array}
\usepackage{color}
\usepackage{float}

\allowdisplaybreaks

\def\->{\rightarrow}

\newtheorem{theorem}{Theorem}
\newtheorem{lemma}{Lemma}
\newtheorem{corollary}{Corollary}

\newtheorem{remark}{Remark}
\newtheorem{example}{Example}
\newtheorem{definition}{Definition}

\begin{document}

\title{Optimal Storage Allocation for Wireless Cloud Caching Systems with
a Limited Sum Storage Capacity}
\author{Bi~Hong,~\IEEEmembership{Student Member,~IEEE} and~Wan~Choi,~\IEEEmembership{Senior Member,~IEEE}
\thanks{This work was supported by ICT R\&D program of MSIP/IITP. [B0114-16-0009, A research on a novel communication system using storage as wireless communication resource]}
\thanks{B.~Hong and W.~Choi are with the School of Electrical Engineering, Korea Advanced Institute of Science and Technology (KAIST), Daejeon, Korea. (E-mail: bi\_hong@kaist.ac.kr, wchoi@kaist.edu)}
\thanks{Parts of this paper were presented in IEEE International Conference on Acoustics, Speech, and Signal Processing (ICASSP), Florence, Italy, May 2014 \cite{ICASSP}.}
} \maketitle

\vspace{-0.8 in}
\begin{abstract}
In wireless cloud storage systems, the recovery failure probability depends on not only wireless channel conditions but also storage size of each distributed storage node. For an efficient utilization of limited storage capacity and the performance characterization of allocation strategies, we asymptotically analyze the recovery failure probability of a wireless cloud storage system with a sum storage capacity constraint for both high SNR regime and low SNR regime. Then, we find the optimal storage allocation strategy across distributed storage nodes in terms of the asymptotic recovery failure probability. Our analysis reveals that
the maximal symmetric allocation is optimal for high SNR regime and the minimal allocation (with $\lfloor T\rfloor$ complete storage nodes and an incomplete storage node) is optimal for low SNR regime, where $T$ is the sum storage capacity.  Based on the numerical investigation, we also show that in intermediate SNR regime, a balance allocation between the minimal allocation and the maximal symmetric allocation would not be required if we select one between them according to SNR.

\end{abstract}

\section{Introduction}

\label{sec:intro} In recent years, the advent of various kinds of social networks, high-definition video streaming, and ubiquitous cloud storage entails large-scale storage in communication networks. Cloud storage systems  are able to meet the demand on large-scale storage capacity only with limited storage capacity of each storage node. Moreover, the cloud storage systems improve reliability of data storage and recovery since they are robust to  failures of individual storage nodes to a certain degree. The robustness also makes repair and maintenance easy when an appropriate network coding technique is adopted.

Theoretically, successful recovery in a cloud storage system is possible if and only if the corresponding max-flow or min-cut from the storage nodes is greater than or equal to the size of the original data object. To implement this feature, given two positive integers $k$ and $n$, a $(n,k)$ maximum distance separable (MDS) code can be used to encode and store the original data into $n$ storage nodes such that recovery of the original data
is possible with $k$ out of $n$ nodes.
For practical implementations, erasure coding is known to be more reliable than duplication of the file  \cite{basic_erasure}. Reed-Solomon code \cite{RS_code} is the most popular one for practical implementation of MDS codes and Reed-Solomon code with information dispersal algorithm (IDA) for distributed storage was investigated in \cite{IDA, Ocean, TotalRecall, Dash}. Fountain codes\cite{Fountain} and low-density parity-check (LDPC) \cite{LDPC} are also known to have approximate MDS properties. Especially, raptor code, the first known class of fountain codes as well as online codes \cite{Online},  is another example of rateless erasure codes and  provides linear time complexity of encoding and decoding.

When a storage node fails, the code repairing problem in a distributed storage system is addressed in \cite{Dimakis:NC_sum} where the code repairing techniques are categorized into exact repair, functional repair, and exact repair of systematic parts. The blocks newly reconstructed by the functional repair preserve the MDS property and enable data recovery, but they are not the same as the original blocks. On the contrary, in the exact repair, the failed blocks are exactly reconstructed. The exact repair of systematic parts is a hybrid repair model standing between the functional repair and the exact repair.
tThe functional repair problem in distributed storage systems was studied in \cite{Dimakis:NC}, interpreting the problem as a multicasting problem over an information flow graph. For the exact repair, it was shown in \cite{Rashmi:MBR} that the optimal minimum bandwidth regenerating (MBR) code can be found for $d=n-1$, where $d$ and $n$ are the number of surviving nodes and the number of storage nodes, respectively. For the exact repair, the exact minimum storage regenerating (MSR) code  based on interference alignment was proposed in  \cite{Suh:MSR}, when $\frac{k}{n}\leq\frac{1}{2}$ and $d\geq2k-1$ where $k$ is the minimum number of nodes required for data
recovery.

The capacity of multicast networks with network coding was given in the pioneering work of Ahlswede et al. \cite{Info_flow}. It was also shown  in \cite{RL_NC} that the random linear network coding over a sufficiently large finite field asymptotically achieved the multicast capacity. For distributed storage, network coding was introduced in \cite{DS_SENSOR1, DS_SENSOR2, DS_SENSOR3} for wireless sensor network. In \cite{DS_SENSOR1, DS_SENSOR2}, decentralized erasure codes inspired by network coding on random bipartite graphs were proposed for distributed multiple sources and their applications to sensor networks were presented. Another linear technique to increases data persistence in wireless sensor networks was proposed and compared to other codes  when the positions and topology of nodes were unknown in \cite{DS_SENSOR3}.  Pyramid codes for flexibly exploiting the tradeoffs between total storage space and access efficiency in a distributed storage system was investigated in \cite{DS_APP1}. In \cite{DS_APP2}, partial network coding (PNC) generalizing network coding was investigated for data collection in distributed sensor networks. For a joint storage and transmission problem, \cite{DS_APP3} showed that a linear coding strategy with file splitting (instead of coding) achieved optimality in total cost including the individual cost of updating, storing and retrieving.
Other key issues on network codes for distributed storage can be referred to \cite{Dimakis:NC_sum}.

For an efficient utilization of limited storage capacity, resource allocation in distributed storage systems has been actively explored. A storage allocation problem under a constraint of total storage capacity was studied in \cite{Ho:DSallo}, where the recovery probability at the data collector was analyzed when the link connections from each node to the data collector are modeled as independent and identically distributed (i.i.d.) Bernoulli random variables with parameter $p$.  It was found in \cite{Ho:DSallo} that the maximal symmetric allocation that equally distributes the total storage capacity to the storage nodes is optimal if the total storage budget is large enough. The gap between the maximal symmetric allocation and the optimal solution vanishes as the total number of storage nodes grows, when $pT>1$, where $T$ is the normalized total storage capacity. If the total storage budget is small, the minimal allocation was shown to be optimal, where the total storage budget is distributed to the minimized number of storage nodes only. In \cite{Dimakis:DSalloH}, these results were extended to a distributed storage system with heterogeneous links where the connection probability from node $i$ to the data collector is $p_i$.

Most of the previous papers on network coding in distributed storage systems rely mainly on simple graph networks with reliable links.
System design and performance analysis of distributed storage systems in fading channels is crucial. For example, diversity gains by multiple antennas or multiple nodes \cite{Tse:Cooperative_diversity,Tse:DMT} in distributed storage systems are required to be properly analyzed and evaluated in fading channels. However, there have been very few studies on distributed storage allocation with non-reliable fading links so far. 
Although some works, such as \cite{Ho:DSallo,Dimakis:DSalloH}, tried to take account of unreliable links with Bernoulli random variables, they failed to exactly account for the key features of wireless links such as channel fading.
For example, diversity gains by multiple antennas or cooperative nodes \cite{Tse:Cooperative_diversity, Tse:DMT} in distributed storage systems are required to be properly analyzed and evaluated in fading channels. Despite the importance of system design and performance analysis of distributed storage systems in fading channels, there have been few studies
on distributed storage allocation with non-reliable fading links so far.

In this context, we consider a wireless cloud storage system where data storage and recovery are carried out through wireless fading channels.
A personal cloud storage system might be a good application of our system model. Contrary to public cloud storage on internet like Dropbox or Google drive, users directly access nearby wireless storage to store or retrieve their data in a personal cloud storage system. Apparently, distributed storage techniques and resource allocation under limited storage space are key techniques enabling personal wireless cloud storage systems. In this regard, the problem of storing and retrieving a file over multiple storage devices under limited sum storage capacity is also a fundamental issue to be addressed, from which we can identify the optimal file portions to store at each storage device. In system design or deployment, the problem reveals the optimal storage size at each storage device under a limited budget of total available storage space.
A cloud edge computing system can also be an extended application. Cloud edge computing includes the processes of distributing input data generated from a user to  nearby computing entities (or devices) and delivering output data from the computing entities to the user. The processes are basically identical with the processes of storing and retrieving data over distributed storage devices. Therefore, our system model can be used as a framework for wireless cloud edge computing system.

In our system model, the recovery failure event occurs either when a data object is not correctly stored at the distributed storage nodes or when the data recovery at the collector fails. We analyze and characterize recovery failure probability in an asymptotic sense for high SNR regime and low SNR regime, and quantify the effect of limited storage capacity on the asymptotic recovery failure probability in high SNR regime.  The contributions of this paper are summarized as follows.

\begin{itemize}
\item We find the \emph{asymptotically} optimal storage allocation under a constraint of total storage capacity for high SNR regime and low SNR regime.

\item Using exponential equality analysis, we show that the optimal allocation strategy for high SNR regime is maximal symmetric spreading of the total storage budget across the distributed storage nodes.

\item We show that for low SNR regime, the minimal allocation with $\lfloor T\rfloor$ complete and 1 incomplete storage nodes is optimal. In the minimal allocation, $\lfloor T \rfloor - T$ storage budget, which is the remaining storage budget after allocating $\lfloor T \rfloor$ budget to $\lfloor T \rfloor$ storage nodes, is not required to be allocated.

\item  In low SNR regime, the storing phase becomes the performance bottleneck and thus the recovery failure probability mainly depends on the cardinality of the decoding set of which elements are the storage nodes that have stored the object successfully.

\item Based on numerical investigation, we show that in intermediate SNR, a balance allocation between the minimal allocation and the maximal symmetric allocation would not be necessary if we properly switch them according to SNR.

\end{itemize}

The rest of the paper is organized as follows. In Section II, we present our system model and notations about the exponential equality. Section III describes how
the wireless cloud storage system operate. The optimal allocation for asymptotic SNR region is investigated and analyzed in Section IV. In Section V, we
present numerical results. Finally, we conclude this paper in Section VI.

\section{System Model and Notations}
\subsection{System and Channel Model}

        \begin{figure}[t!]
        \begin{center}
        \includegraphics[scale=0.75]{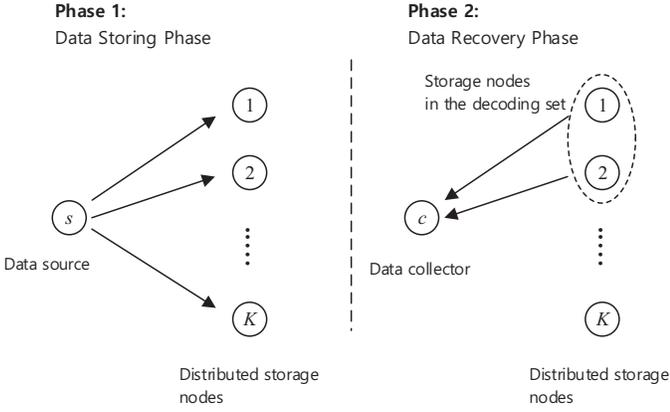}
        \caption{A wireless distributed storage system model constituted by data storing and recovery phases \label{fig:system}}
        \end{center}
        \end{figure}
As shown in Fig. \ref{fig:system}, the number of storage nodes is set to $K$. Data source, data collector, and storage nodes are denoted by $s$, $c$ and $\{1,\ldots,K\}$, respectively. The collector node recovers the stored data from the storage nodes. If we consider the scenario when a mobile user stores its data on cloud storage, the collector node will be the same as the data source, so the direct link from the source node to the data collector does not exist. The channel gain from node $i$ to node $j$ is denoted by $h_{i,j}$  and follows the complex Gaussian distribution with zero mean and unit variance, i.e., $h_{i,j}\sim\mathcal{CN}(0,1)$.
For simplicity, we assume all channels  independent and identically distributed (i.i.d.) and the path loss are ignored, but the result of this paper can be extended to a general case with ease.
To facilitate tractable analysis without losing key insights on system design, we assume i.i.d. channels. Moreover, if storage nodes are not close one another in a rich scattering environment, the independent channel model can be effective.The analytic framework developed in i.i.d. channels will be also useful in non i.i.d. channels.
For file storing and recovery, a file is divided into many data blocks and the storing and recovery processes are carried out for each data block. Because  mobility of both storage devices and user is limited in personal cloud storage systems, the Doppler spread is not likely to be large. For example, in low mobility environments (i.e, moving speed is less than 5 m/sec), the coherent time becomes approximately 20 ms with 2.4GHz carrier frequency. On the other hand, symbol duration  is on the order of tens of microseconds and the  supported data rate reaches more than 1 Gbps in current wireless communication protocols such as Wi-Fi direct. Consequently, each data block size can be as large as up to 2.5 Mbytes under the coherent time requirement. This block size is large enough to implement ideal MDS coding and given the fact that storage devices like HDD or SSD use block size of 4 KB -- 2,048 KB, the requirement of coherence time can be readily met for each data block. Therefore, we assume that the coherence time is longer than each storage period and recovery period  for each data block 
and thus the channel gains do not change during each period but the channel gains independently change between the periods.

An additive white Gaussian noise (AWGN) is denoted by $z_{i}$ and follows the complex Gaussian distribution with zero mean and variance $N_0$, i.e., $z_{i}\sim\mathcal{CN}(0,N_0)$. The average signal-to-noise ratio (SNR) of a link is denoted as $\rho=P/N_0$ and the amount of coded data stored in storage node $i\in\{1,\ldots,K\}$ is denoted as $a_i$.

In our model, each storage node does not have an individual restriction on the storage size but the maximum amount of data stored at each storage node is $T$ because the total storage budget is limited to $T$ such that
$\sum_{i=1}^{K}a_i \leq T$.
We assume that the size of data object is normalized to be unit compared to the total storage capacity $T$ as in other literature \cite{Ho:DSallo}, which simplifies design and analysis of distributed storage systems. 
A data object  corresponds to a divided data block and 
the recovery probability to be analyzed is for each data block, not for the whole file.
Total storage constraint $T$ and allocated budget $a_i$ is also identically used for a data block consisting the file because they are already normalized values compared to the original data. In other words, $a_i$ can also denotes the amount of MDS coded fraction stored in storage node $i$.

\subsection{Storage and Recovery Operation}
As shown in Fig. \ref{fig:system}, the storage and recovery operation is constituted by two basic phases. The first one is for storing data object in the distributed storage nodes and the second one is for recovering the stored data. 
These two phases are decoupled in time since the stored data are recovered later although the first phase affects the data recovery in the second phase. That is, the storage phase and the recovery phase are not  concurrently entered into.

\begin{enumerate}
   \item Storage Phase: The source node broadcasts a data object to the cloud storage nodes to store the data object during a given time period. If a storage node has successfully decoded the broadcast data object from the source, the storage node converts the decoded data object into suitable MDS coded blocks as much as its allocated storage size $a_i$, and stores it.  Storage node $i$ successfully decodes the data object from the source node only when the instantaneous mutual information between the source node and storage node $i$ is greater than or equal to the required rate of the data object, i.e., $\log_2 (1 + \rho |h_{s,i}|^2) \geq Q. $  The value of threshold $Q$ depends on the adopted modulation and coding scheme (MCS) and bandwidth for data transmission. For example, a typical range of required SNR in 802.11n Wi-Fi is around from 10 to 30 dB \cite{RX_sensitivity}. Taking account of specific system parameters of the adopted wireless communication protocol, the threshold $Q$ corresponding to a required SNR value can be computed.
The amount of MDS coded data stored at node $i\: (\in \mathcal{D})$ is equal to its allocated storage size $a_i$, where $\mathcal{D}$ is the decoding set whose elements are the storage nodes which have successfully decoded the broadcast data object.  If $\sum_{i\in \mathcal{D}}a_i < 1$, the data object is not properly stored  because the data object cannot be properly recovered from the stored MDS coded data \cite{Ho:DSallo}. On the other hand, if $\sum_{i\in \mathcal{D}}a_i \geq 1$, the collector has a chance to properly recover the data object from the stored MDS coded data, depending on channel conditions from the storage nodes to the collector. Note that instead of MDS coding, random linear coding over a sufficiently large field can be used for data object recovery \cite{RL_NC}.

\item Recovery Phase: When the data collector wants to recover the data object, it requests the storage nodes in the decoding set to send the MDS coded blocks of the stored data. The storage phase and the recovery phase are assumed to have the same time length for simplicity. Let $\hat{i}\in\{1,2,\ldots,|\mathcal{D}|\}$ be the index of the storage node in the decoding set. They are indexed in descending order of the channel power gain from the storage node to the collector such that
$|h_{\hat{1},c}|^2\geq|h_{\hat{2},c}|^2\geq\cdots\geq |h_{\hat{|\mathcal{D}|},c}|^2$.
Then, storage node $\hat{i}$ transmits during $t_{\hat{i}}$ fraction of the time period allocated for the recovery phase, where $t_{\hat{i}}$ varies with the cardinality of the decoding set and has to satisfy $t_{\hat{i}} \leq a_{\hat{i}}$ and $\sum_{i=1}^{|\mathcal{D}|}t_{\hat{i}}=1$ which is a constraint from the property of MDS code. The data collector is assumed to know the index and the stored data size of each storage node by proper signaling with the storage nodes. With the help of pilot symbols from each storage node in the decoding set, the data collector estimates the channel gains from the storage node.  To focus on developing the optimal distributed caching strategy, we assume perfect channel state information at the data collector and discard the signaling overhead. Based on the channel gains, the data collector determines the time fractions, $\{t_i\}$, by solving the following optimization problem:
\begin{align}
  &\underset{t_{\hat{i}}}{\arg\max}\sum_{i=1}^{|\mathcal{D}|}
  t_{\hat{i}}\log(1+|h_{{\hat{i}},c}|^2\rho)\quad\\
  &\text{subject to } \sum_{i=1}^{|\mathcal{D}|} t_{\hat{i}} =1,
    \quad t_{\hat{i}}\leq a_{\hat{i}}.\label{eq:first_convex}
\end{align}
The optimized value of $t_{\hat{i}}^*$ is given by
\begin{align}
&t_{\hat{i}}^*=a_{\hat{i}}\cdot\textbf{1}\left[\Sigma_{\hat{j}=1}^{i}a_{\hat{j}}<1\right]\nonumber\\
&+\left(1-\Sigma_{\hat{j}=1}^{i-1}a_{\hat{j}}\right)\cdot\textbf{1}\left[\Sigma_{\hat{j}=1}^{i}a_{\hat{j}}>1 \text{ and } \Sigma_{\hat{j}=1}^{i-1}a_{\hat{j}}\leq1 \right]
\end{align}
where $\mathbf{1}(\cdot)$ is the indicator function which returns 1 if the argument is true or 0 otherwise, which suggests that longer transmission time is allocated to the storage node with a stronger channel gain under the  storage constraint and the transmit time constraint.

According to the determined $\{t_i\}$, the data collector receives the stored MDS coded block from storage node $\hat{1}$ during the $t_{\hat{1}}$ time portion. Then, it moves on to storage node $\hat{2}$ and receives the stored MDS coded block during  $t_{\hat{2}}$. In this way, the data collector receives the stored MDS coded date from the storage nodes in the decoding set. The data object is successfully recovered if
\begin{align}
  \sum_{i=1}^{|\mathcal{D}|}t_{\hat{i}}\log(1+|h_{\hat{i},c}|^2\rho)>Q.
\end{align}
Otherwise, recovery of the stored data object fails.
 \end{enumerate}

\subsection{Notations}
The exponential equality is denoted as the symbol $\doteq$, i.e.,
$f(\rho)\doteq\rho^{b}$, when
$
\lim_{\rho\->\infty}\frac{\log\left(f(\rho)\right)}{\log(\rho)}=b
$
where $b$ is called the \emph{exponential order} of
$f(\rho)$. The exponential inequalities denoted by $\dot{\leq}$ and
$\dot{\geq}$ are similarly defined. $(x)^+$ is used to denote
$\max\{x,0\}$. $\mathbb{R}^N$ is the set of real $N$-tuples, while
$\mathbb{R}^{N+}$ denotes the set of nonnegative real $N$-tuples.
For any set $\mathcal{O}\in\mathbb{R}^N$, the intersection of the
set and $\mathbb{R}^+$ is denoted by $\mathcal{O}^+$, i.e.,
$\mathcal{O}^+=\mathcal{O}\cap\mathbb{R}^{N+}$. Assume that $h$ is a
Gaussian random variable with zero mean and unit variance. Then, the
asymptotic probability density function (pdf) of the exponential
order of $1/|h|^2$ denoted by $v$ is obtained as
$
p_v=\lim_{\rho\->\infty}\ln(\rho)\rho^{-v}\exp(-\rho^{-v})\quad\text{where}\quad
v=-\lim_{\rho\->\infty}\frac{\log(|h|^2)}{\log(\rho)}.\label{eq:exppdf}
$
By limiting $\rho$ to infinity, the pdf in \eqref{eq:exppdf} is given
by
$
p_v\doteq
\rho^{-\infty}=0~\textrm{for } v< 0,\quad
p_v\doteq
\rho^{-v}~\textrm{for }v \geq 0.
$
Thus, for independent random variables $\{v_j\}_{j=1}^{K}$
distributed identically to $v$, the probability $P_{\mathcal{O}}$
that $(v_1,\ldots,v_K)$ belongs to set ${\mathcal{O}}$
 can be characterized by
{\fontsize{11}{11}
\begin{align}
P_{\mathcal{O}}\doteq \rho^{-d_0}, \quad\text{for }\quad
d_0=\inf_{(v_1,\ldots,v_K)\in
{\mathcal{O}}^+}\sum_{j=1}^{K}v_j\label{eq:expresult}
\end{align}
}%
provided that ${\mathcal{O}}^+$ should be non-empty. In other words,
the exponential order of $P_{\mathcal{O}}$ depends on
${\mathcal{O}}^+$ only.
The list of symbols used in the paper is given in Table \ref{table:syms}.
\begin{table}[t]
\caption{List of symbols and Its descriptions}
    \begin{center}
         \begin{tabular}{  c | l }
         \hline
         $h_{i,j}$ & channel coefficient from $i$ to $j$  \\ \hline $\rho$ & average received SNR \\ \hline
         $a_i$ 	& individual storage constraint at storage $i$  \\ \hline $K$ & the number of storage nodes \\ \hline
         $\mathbf{a}$ 	& storage allocation vector  \\ \hline $s$ & source\\ \hline
         $T$ & total storage capacity  \\ \hline $c$ & data collector\\ \hline
         $Q$ & accumulated rate threshold  \\ \hline $d(\cdot)$ & exponential order\\ \hline
         \end{tabular}
    \end{center}\label{table:syms}
\end{table}	
\section{Asymptotic Analysis of Recovery Failure Probability}

For the system model described in the previous section, the
recovery failure probability (i.e., the complimentary recovery probability) is hard
to obtain in  closed form, as noted in \cite{Ho:DSallo}, even if interesting wireless ingredients are not incorporated. In this section, we instead explore the optimal storage allocation in an
asymptotic sense. Exponential order determines the decreasing speed of recovery errors in high SNR regime. Consequently, large exponential order offers low recovery failure probability if SNR is sufficiently high. On the other hand, the recovery failure probability is not characterized well by exponential order in low SNR regime, so the dominant order of recovery probability helps understand recovery performance in low SNR regime.

\subsection{Optimal Storage Allocation in High SNR Regime}
We analyze the recovery failure probability for high SNR regime in this subsection to understand
its asymptotic behavior. That is, we derive the exponential order of
the recovery failure probability and find the optimal storage
allocation to maximize the exponential order. The exponential order
characterizes the decreasing tendency of the recovery failure
probability versus SNR, and is interpreted as diversity order if bit
error probability or outage probability is considered in conventional
wireless communication systems. Contrary to the conventional
diversity order, the exponential order of the recovery failure
probability is determined by not only the number of independent
fading paths for data storage and recovery but also the limited total
storage capacity.

\begin{lemma}
For given storage allocation $\mathbf{a}=\{a_1,\ldots,a_K\}$,
the exponential order of the recovery failure probability is
lower and upper bounded, respectively, by
\begin{align}
  d(\mathbf{a}) &\geq \min_{\mathcal{D}\subseteq\{1,\ldots,K\}}\left( K-|\mathcal{D}|+\min_{i\in\mathcal{D}}{t_i}^{-1}\cdot\mathbf{1}[\sum_{i\in\mathcal{D}}a_i\geq1]\right)\label{thm:d_order1}\\
  \text{ and }&d(\mathbf{a}) \leq \min_{\mathcal{D}\subseteq\{1,\ldots,K\}}\left( K-|\mathcal{D}|+|\mathcal{D}|\cdot\mathbf{1}[\sum_{i\in\mathcal{D}}a_i\geq1]\right).\label{thm:d_order2}
  \end{align}
  \label{lem:d_order}
\end{lemma}
\begin{IEEEproof}
Refer to Appendix \ref{app:1}.  
\end{IEEEproof}

\begin{theorem}
  Under a total storage capacity $T\:(>1)$ constraint, the optimal
  storage allocation in terms of exponential order is to
  maximally and symmetrically allocate the sum storage capacity
  across all storage nodes.
  \label{thm:allo}
\end{theorem}
\begin{IEEEproof}
Our optimization problem is formulated as
  \begin{align}
    &\max_{\mathbf{a}} ~d(\mathbf{a}) \nonumber\\
    \text{subject to } & a_1+a_2+\cdots+a_K=T,\quad a_i\geq t_i\geq 0, \quad\forall i.
  \end{align}
For an arbitrary storage allocation $\mathbf{a}$, the upper bound of $d(\mathbf{a})$ in \eqref{thm:d_order2} is determined by the maximum value of the cardinality $|\mathcal{D}|$  of a decoding set  satisfying
$\sum_{k\in\mathcal{D}}a_k<1$. Let this decoding set be $\mathcal{D}_{\text{UP}}(\mathbf{a})$. Obviously, $\mathcal{D}_{\text{UP}}(\mathbf{a})$ is a set of the nodes whose allocated storage sizes are the
 $|\mathcal{D}_{\text{UP}}(\mathbf{a})|$ smallest ones, i.e.,
 $\mathcal{D}_{\text{UP}}(\mathbf{a})=\{a_1^{\uparrow},a_2^{\uparrow},\ldots,a_{|\mathcal{D}_{\text{UP}}(\mathbf{a})|}^{\uparrow}\}$
where $a_i^{\uparrow}$ denotes the $i$-th smallest storage size in $\mathbf{a}$, and  $|\mathcal{D}_{\text{UP}}(\mathbf{a})|$  is determined such that
 $\sum_{i=1}^{|\mathcal{D}_{\text{UP}}(\mathbf{a})|}a_i^{\uparrow}<1\leq\sum_{i=1}^{|\mathcal{D}_{\text{UP}}(\mathbf{a})|+1}a_i^{\uparrow}$.

To maximize the upper bound of $d(\mathbf{a})$, we have to find an allocation which yields the smallest $|\mathcal{D}_{\text{UP}}(\mathbf{a})|$.
Consider the symmetric storage allocation $\mathbf{a}_{\text{sym}}$ in which the allocated storage sizes are the same as $\frac{T}{K}$. Then, for the symmetric allocation, the following inequalities hold.
\begin{align}\label{ineq1}
  \frac{\sum_{i=1}^{|\mathcal{D}_{\text{UP}}(\mathbf{a}_{\text{sym}})|}a_i^{\uparrow}}{|\mathcal{D}_{\text{UP}}(\mathbf{a}_{\text{sym}})|}\leq\frac{T}{K},
 \end{align}
 \begin{align}
  \quad\frac{T}{K}\cdot|\mathcal{D}_{\text{UP}}(\mathbf{a}_{\text{sym}})|<1\label{ineq2}
\end{align} where \eqref{ineq1} is due to the fact that an average with the $|\mathcal{D}_{\text{UP}}(\mathbf{a}_{\text{sym}})|$  smallest ones is less than an arithmetic average;  \eqref{ineq2} is because $|\mathcal{D}_{\text{UP}}(\mathbf{a}_{\text{sym}})|$ is determined to satisfy
$\sum_{k\in\mathcal{D}}a_k<1$ for the symmetric allocation.
Combining \eqref{ineq1} and \eqref{ineq2}, we have $\sum_{i=1}^{|\mathcal{D}_{\text{UP}}(\mathbf{a}_{\text{sym}})|}a_i^{\uparrow}<1$. Since $\sum_{i=1}^{|\mathcal{D}_{\text{UP}}(\mathbf{a})|}a_i^{\uparrow}<1\leq\sum_{i=1}^{|\mathcal{D}_{\text{UP}}(\mathbf{a})|+1}a_i^{\uparrow}$, there exists a set $\mathcal{D}_{\text{UP}}(\mathbf{a})$ such that $\{a_1^{\uparrow},a_2^{\uparrow},\ldots,a_{|\mathcal{D}_{\text{UP}}(\mathbf{a}_{\text{sym}})|}^{\uparrow}\}\subseteq\mathcal{D}_{\text{UP}}(\mathbf{a})$, which implies that $|\mathcal{D}_{\text{UP}}(\mathbf{a}_{\text{sym}})|\leq|\mathcal{D}_{\text{UP}}(\mathbf{a})|$. Therefore, the allocation corresponding to the minimum
 $|\mathcal{D}_{\text{UP}}(\mathbf{a})|$ is the symmetric storage allocation,  $\mathbf{a}_{\text{sym}}$.

With the symmetric allocation, the lower bound of $d(\mathbf{a})$ coincides with the maximized upper bound. That is,  the term $\min_{i\in\mathcal{D}}t_i^{-1}$ in lower bound \eqref{thm:d_order1} becomes $|\mathcal{D}|$ with  the symmetric allocation. Since the maximum values of the upper and lower bounds coincide, we conclude that the symmetric allocation maximizes the exponential order.
\end{IEEEproof}

\begin{corollary}\label{col:opt_d}
  For given  $K$ and  $T$, the optimal exponential order of the recovery failure probability is
  \begin{align}\label{eq:opt_d}
    d^*(K,T)=K-\left\lceil\frac{K}{T}\right\rceil+1
  \end{align}
  with the optimal storage allocation policy.
\end{corollary}
\begin{IEEEproof}
The proof of Theorem \ref{thm:allo} showed that for symmetric storage allocation, $|\mathcal{D}_{\text{UP}}(\mathbf{a_{\text{sym}}})|T/K<1$. Therefore, the maximum possible cardinality of the decoding set is  $|\mathcal{D}_{\text{UP}}(\mathbf{a_{\text{sym}}})|=\lceil K/T\rceil-1$. Plugging this in Lemma \ref{lem:d_order}, we obtain \eqref{eq:opt_d}.
\end{IEEEproof}

\begin{remark}
Theorem \ref{thm:allo} is on the same line with the result of
\cite{Ho:DSallo}. Theorem \ref{thm:allo} exhibits that the maximal symmetric spreading of the sum storage capacity
$T \:(>1)$ yields the optimal recovery probability in terms of exponential order even for
wireless distributed storage systems suffering from channel fading.
 \label{remark:1}
\end{remark}
\begin{remark}
The optimal exponential order of the recovery failure probability is
bounded above and below by $ \left(1-\frac{1}{T}\right)K \leq d^*(K,T) \leq
\left(1-\frac{1}{T}\right)K+1. $ Thus, the approximated slope of the exponential
order is $\left(1-\frac{1}{T}\right)$, which is strictly less than
1, for the sum storage capacity $T$.
\label{remark:2}
\end{remark}

Although the exponential order well characterizes asymptotic
behavior of the recovery failure probability, we also derive a
high SNR approximation of the recovery failure probability for concrete
understanding of recovery success and failure in high SNR regime, when the sum storage capacity $T$ is maximally and symmetrically spread to the storage nodes.

\begin{theorem}
When SNR is sufficiently high, the recovery failure probability is
approximated as
  \begin{align}
  &\text{Pr}_{f}^{\text{high}}[Q] \approx \binom{K}{\lceil \frac{K}{T}\rceil-1}(2^Q-1)^{K-\lceil \frac{K}{T}\rceil+1}\rho^{-(K-\lceil \frac{K}{T}\rceil+1)}
  \end{align}
\end{theorem}

\begin{IEEEproof}
Lemma \ref{lem:d_order} and Theorem \ref{thm:allo} indicate that
when SNR is sufficiently high, the exponential order of the recovery failure probability is
dominated by $|\mathcal{D}_{\text{UP}}(\mathbf{a}_{\text{sym}})|$.
For $\mathbf{a}_{\text{sym}}$ and $|\mathcal{D}_{\text{UP}}(\mathbf{a}_{\text{sym}})|$, the recovery failure probability when SNR is sufficiently high is obtained as

\begin{align}
  \text{Pr}_{f}[Q]&=\sum_{\mathcal{D}\subseteq\{1,\ldots,K\}}\text{Pr}_{f}[Q\:|\:\mathcal{D}]\text{Pr}[\mathcal{D}]\\
  &\overset{(a)}{\approx}\binom{K}{\lceil \frac{K}{T}\rceil-1}\text{Pr}\left[|\mathcal{D}|=\left\lceil K/T\right\rceil-1\right]\nonumber\\
  &=\binom{K}{\lceil \frac{K}{T}\rceil-1}\text{Pr}[\log_2(1+\rho|h|^2)>Q]^{\lceil \frac{K}{T}\rceil-1}\nonumber\\
  &\times\text{Pr}[\log_2(1+\rho|h|^2)<Q]^{K-\lceil \frac{K}{T}\rceil+1}\\
  &=\binom{K}{\lceil \frac{K}{T}\rceil-1}\left(\exp\left(\frac{2^Q-1}{\rho}\right)\right)^{\lceil\frac{K}{T}\rceil-1}\nonumber\\
  &\times\left(1-\exp\left(\frac{2^Q-1}{\rho}\right)\right)^{K- \lceil \frac{K}{T}\rceil+1}\\
  &\overset{(b)}{\approx}\binom{K}{\lceil \frac{K}{T}\rceil-1}\left(\frac{2^Q-1}{\rho}\right)^{K-\lceil \frac{K}{T}\rceil+1}\label{eq:taylor}
\end{align}

where $(a)$ is from the result of Lemma \ref{lem:d_order} with high SNR assumption and $(b)$ is due to Taylor's expansion as $\rho\rightarrow\infty$.
\end{IEEEproof}

\subsection{Optimal Storage Allocation in Low SNR Regime}

Although we have derived the optimal storage capacity allocation for high SNR, it is not clear whether the  derived solution is always optimal for all other SNR regimes.  As stated earlier, the exact closed form expression of the recovery failure probability is hard to obtain due to its mathematical intractability. Thereby, in this subsection, we explore the optimal storage allocation strategy in low SNR regime. To this end, we have to first understand the relationship between the decoding set cardinality and the recovery failure probability. We start with the following definition of a complete storage node.

\begin{definition}
  Storage node $i$ is defined as a complete storage node if it can store a complete data object and the data object can be perfectly recovered from it without help of any other storage nodes. That is, if $a_i\geq1$, storage node $i$ is a complete storage node.
 \end{definition}

\begin{lemma}\label{lemma:1}
Any storage allocation strategy without complete storage nodes has higher recovery failure probability in low SNR than a storage allocation strategy with only one complete storage.
\end{lemma}

\begin{IEEEproof}
Let us consider the following storage allocation strategy:
\begin{align}
  &\mathbf{a}(\epsilon)=\{1-\epsilon,1-\epsilon,\ldots,1-\epsilon\}\nonumber \\
  &\text{ where $\epsilon$ is an arbitrarily small positive value ($\epsilon >0$).}
\end{align}
To recover the data object, at least two storage nodes are required for the allocation $\mathbf{a}(\epsilon)$ and the corresponding recovery probability becomes
\begin{align}
  &\bar{P}_O(\mathbf{a}(\epsilon))
  =\sum_{k=2}^{K}\binom{K}{k}\text{Pr[$|\mathcal{D}|=k$]}\nonumber\\
  &\times\text{Pr[recovery from $k$ incomplete storage nodes]}\\
  &=\sum_{k=2}^{K}\binom{K}{k}\text{Pr[$|\mathcal{D}|=k$]}\cdot\text{Pr$\Big[\:\sum_{i\in\mathcal{D}}t_i\log(1+|h_{i,c}|^2)>Q\:\Big]$}\nonumber\\
  &\overset{(a)}{\leq}\sum_{k=2}^{K}\binom{K}{k}\text{Pr[$|\mathcal{D}|=k$]}\cdot\text{Pr$\Big[\:\max_{i\in\mathcal{D}}\log(1+|h_{i,c}|^2)>Q\:\Big]$}\label{eq:(a)}\\
  &=\sum_{k=2}^{K}\binom{K}{k}e^{-k\cdot\frac{2^Q-1}{\rho}}\left(1-e^{-\frac{2^Q-1}{\rho}}\right)^{K-k}\nonumber\\
  &\times\left(1-\left(1-e^{-\frac{2^Q-1}{\rho}}\right)^k\right)\\
  &=e^{-2\cdot\frac{2^Q-1}{\rho}}\sum_{k=2}^{K}\binom{K}{k}e^{-(k-2)\cdot\frac{2^Q-1}{\rho}}\nonumber\\
  &\times\left(1-e^{-\frac{2^Q-1}{\rho}}\right)^{K-k}\left(1-\left(1-e^{-\frac{2^Q-1}{\rho}}\right)^k\right)\\
  &\overset{(b)}{\leq}e^{-2\cdot\frac{2^Q-1}{\rho}}
\end{align}
where
$(a)$ is from the selection upper bound (i.e., selecting the node with the strongest channel to the data collector among $k$ complete nodes), and $(b)$ is  because as $\rho$ increases,
the rest term except $e^{-2\cdot\frac{2^Q-1}{\rho}}$ goes to 0 and strictly less than 1.
On the other hand, the recovery probability for the storage allocation $\mathbf{a}_1=(1, 0,\ldots,0)$ is given by
\begin{align}
  &\bar{P}_O(\mathbf{a}_1)=\text{Pr[node 1 is in $\mathcal{D}$]}\cdot\text{Pr[recovery from node 1]}\\
  &=\text{Pr[$\log(1+|h_{s,1}|^2\rho)>Q$] $\cdot$ Pr[$\log(1+|h_{1,c}|^2\rho)>Q$]}\\
  &=\bar{P}_O(\mathbf{a}_1) \left(=\exp\left(-\frac{2^Q-1}{\rho}\right)\exp\left(-\frac{2^Q-1}{\rho}\right)\right)
\end{align}
 Consequently, Lemma 1 is proved because $\bar{P}_O(\mathbf{a}(\epsilon))\leq\bar{P}_O(\mathbf{a}_1)$ in low SNR regime means $\mathbf{a}_1$ always shows lower recovery failure probability than any other allocation without complete storage nodes.
\end{IEEEproof}

\begin{lemma}
For any allocation strategy with $K_1$ complete storage nodes and $K_2$ incomplete storage nodes, the recovery failure probability in low SNR regime is higher than an allocation with $K_1+1$ complete storage nodes.  \label{lemma:2}
\end{lemma}
\begin{IEEEproof}
  Let an allocation strategy with $K_1$ complete storage nodes and $K_2$ incomplete storage nodes be $\mathbf{a}_{(K_1,K_2)}$ and an allocation strategy with $K_1+1$ complete storage nodes $\mathbf{a}_{K_1+1}$. Similar to the proof of Lemma 1, we can easily show the recovery probability of $\mathbf{a}_{(K_1,K_2)}$ can be upper bounded as follows:
  \begin{align}
    &\bar{P}_O(\mathbf{a}_{(K_1,K_2)})\nonumber\\
    &=\sum_{k_1=0}^{K_1}\sum_{k_2=0}^{K_2}\binom{K_1}{k_1}\binom{K_2}{k_2}\text{Pr[$|\mathcal{D}|=k_1+k_2$]}\nonumber\\
    &\times\text{Pr[recovery from $k_1+k_2$ storage nodes]}\\
    &\overset{(a)}{\leq}\sum_{k_1=1}^{K_1}\sum_{k_2=0}^{K_2}\binom{K_1}{k_1}\binom{K_2}{k_2}e^{-(k_1+k_2)\cdot\frac{2^Q-1}{\rho}}\nonumber\\
    &\times\left(1-e^{-\frac{2^Q-1}{\rho}}\right)^{K_1+K_2-k_1-k_2}
    \left(1-\left(1-e^{-\frac{2^Q-1}{\rho}}\right)^{k_1+k_2}\right)\nonumber\\
    &+\sum_{k_2=2}^{K_2}\binom{K_2}{k_2}e^{-k_2\cdot\frac{2^Q-1}{\rho}}\left(1-e^{-\frac{2^Q-1}{\rho}}\right)^{K_2-k_2}\nonumber\\
 &\times\left(1-\left(1-e^{-\frac{2^Q-1}{\rho}}\right)^{k_2}\right)\nonumber\\
    &=\binom{K_1}{1}\exp\left(-2\cdot\frac{2^Q-1}{\rho}\right)\left(1-\exp\left(-\frac{2^Q-1}{\rho}\right)\right)\nonumber\\
    &+o\left(\exp\left(-2\cdot\frac{2^Q-1}{\rho}\right)\right)\\
    &\lesssim\:K_1\exp\left(-2\cdot\frac{2^Q-1}{\rho}\right)
  \end{align}
  where $(a)$ is obtained from the selection upper bound as in \eqref{eq:(a)} and $\lesssim$ is the asymptotic inequality which denotes that the inequality is valid for high SNR regime. Because
  \begin{align}
    &\bar{P}_O(\mathbf{a}_{(K_1,K_2)})\lesssim K_1\exp\left(-2\cdot\frac{2^Q-1}{\rho}\right)\nonumber\\
    &<(K_1+1)\exp\left(-2\cdot\frac{2^Q-1}{\rho}\right)\approx \bar{P}_O(\mathbf{a}_{K_1+1}),
  \end{align} $  \bar{P}_O(\mathbf{a}_{(K_1,K_2)}) < \bar{P}_O(\mathbf{a}_{K_1+1})$ holds for $\rho\->\infty$.
\end{IEEEproof}

\begin{lemma}
When the sum storage buget $T$ is allocated to $\lfloor T\rfloor$ complete storage nodes, the remaining $T-\lfloor T\rfloor$ storage budget should be allocated to only one storage node, to achieve the optimal performance in low SNR regime.\label{lemma:3}
\end{lemma}
\begin{IEEEproof}
Refer to Appendix \ref{app:2}.
\end{IEEEproof}

\begin{theorem}
  The optimal storage budget allocation in low SNR regime consists of $\lfloor T \rfloor$  complete storage nodes and one incomplete storage whose storage size is $T-\lfloor T \rfloor$.
\end{theorem}
\begin{IEEEproof}
  The proof is straightforward from Lemmas \ref{lem:d_order}--\ref{lemma:3}.
\end{IEEEproof}

\begin{example}
In low SNR regime, if the sum storage budget for $K=6$ storage nodes is given by $T=2.25$, two nodes should have storage size $a_i=1$ and one node should have storage size $0.25$. Note that this allocation is completely different.
\end{example}

\begin{remark}
When the total storage budget is given by an integer value i.e., $T = \lfloor T \rfloor$, with the optimal storage allocation in low SNR, the number of storage nodes with non-zero allocated memory is $T$ and they are all complete storage nodes. Then,
contrary to the maximal symmetric allocation in high SNR, node selection for selection diversity in the recovery phase is possible. In this case, the  recovery failure probability can be obtained exactly in closed form as in the following corollary.
\end{remark}

\begin{corollary}
  When the total storage budget is an integer value, the recovery failure probability with the minimal allocation is obtained as \begin{align}
    \text{Pr}_f[Q]=\sum_{k=0}^{T}\binom{T}{k}\left(e^{-\frac{2^Q-1}{\rho}}\right)^k
    \left(1-e^{-\frac{2^Q-1}{\rho}}\right)^{T}.
  \end{align}
\end{corollary}
\begin{IEEEproof}
\begin{align}
  &\text{Pr}_f[Q]=\sum_{k=0}^{T}\text{Pr}[|\mathcal{D}|=k]\cdot\text{Pr}_f[Q\:\big|\:|\mathcal{D}|=k]]\nonumber\\
    &=\sum_{k=0}^{T}\prod_{i\in\mathcal{D}}\text{Pr}\left[\log(1+\rho|h_i|^2)>Q\right]\nonumber\\
    &\qquad\times\prod_{j\notin\mathcal{D}}\text{Pr}\left[\log(1+\rho|h_j|^2)<Q\right]\nonumber\\
    &\qquad\times\text{Pr}\left[\log(1+\rho\max_{i\in\mathcal{D}}|h_i|^2)<Q\right]\\
    &=\sum_{k=0}^{T}\binom{T}{k}\exp\left(-\frac{2^Q-1}{\rho}\right)^k
    \left(1-\exp\left(-\frac{2^Q-1}{\rho}\right)\right)^{T-k}\nonumber\\
    &\qquad\times\left(1-\exp\left(-\frac{2^Q-1}{\rho}\right)\right)^k
\end{align}
\end{IEEEproof}
\section{Numerical Result}
\subsection{High SNR Regime}

\begin{figure}
    \centering   
    \subfigure[By the numbers of storage nodes $K$ for $T=2, 5, 12$.]
    {
        \includegraphics[width=0.468\textwidth]{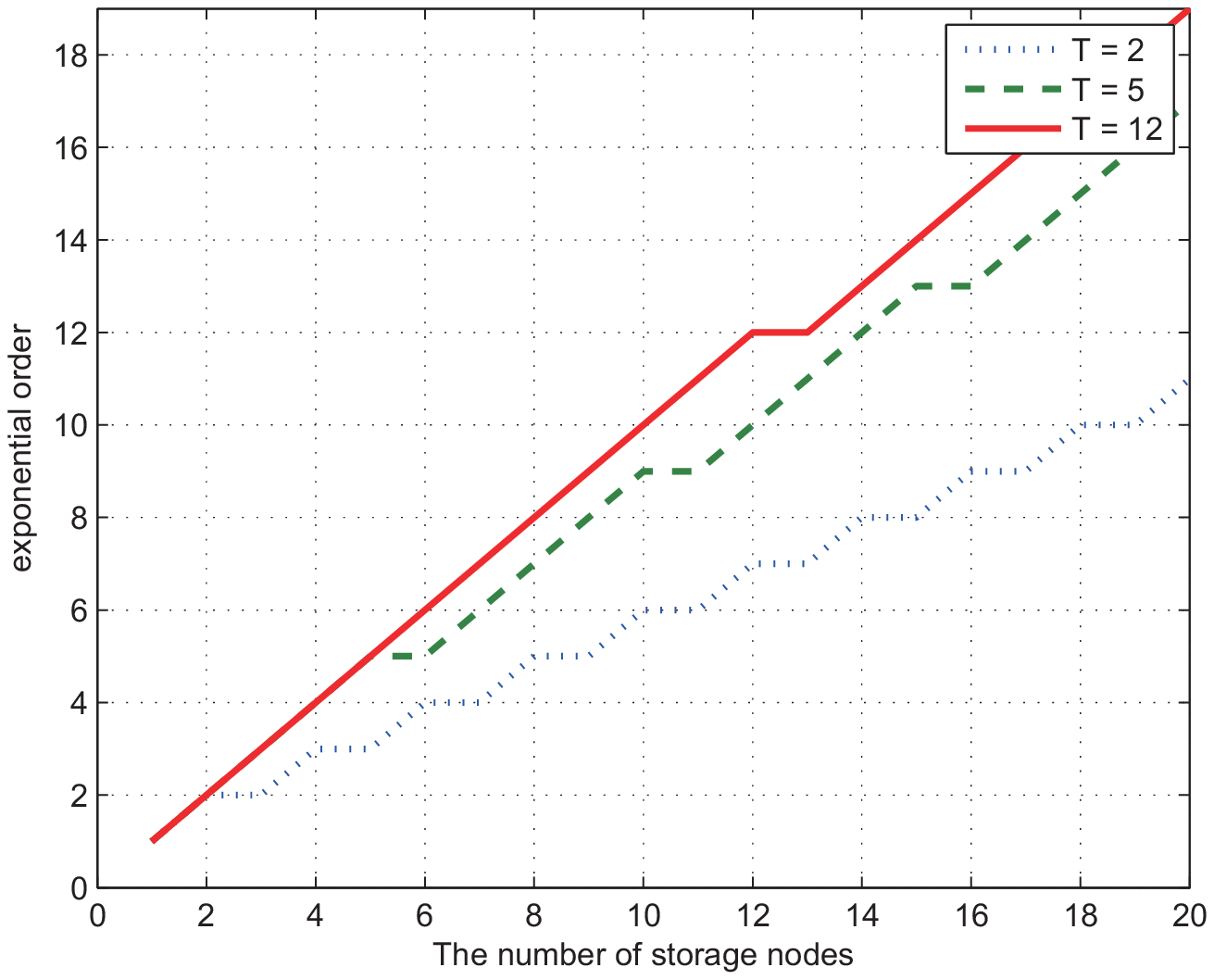}
        \label{fig:sim1}
    }
    \subfigure[By sum storage capacity $T$ for $K=10, 200$.]
    {
        \includegraphics[width=0.468\textwidth]{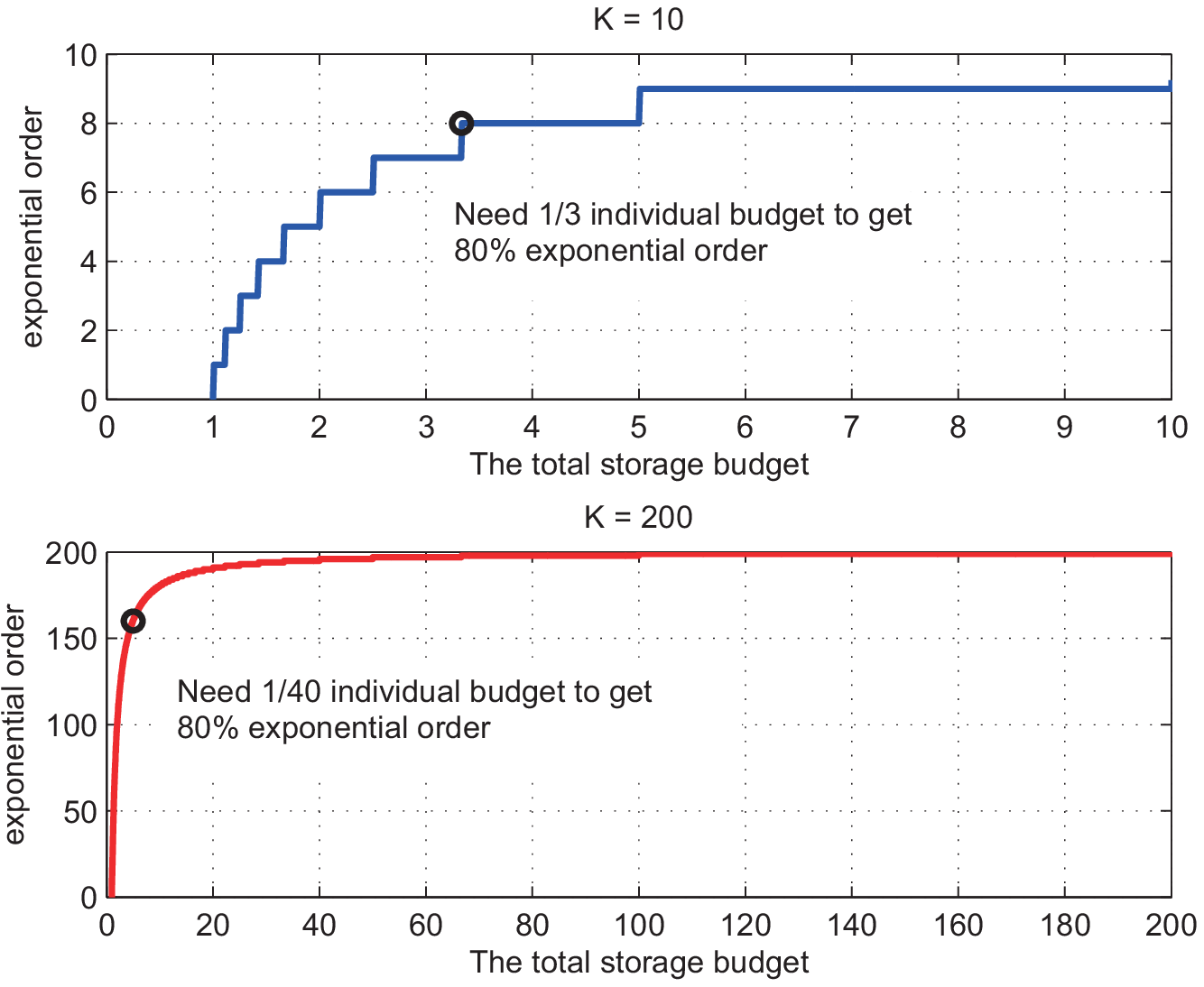}
        \label{fig:sim2}
    }
    \caption{Exponential order growth}
    \label{fig:exp_growth}
\end{figure}

To focus on channel fading effects only, we assume simple topology where the distances from data source/collect to storage nodes are the same. The channel gains of each link are i.i.d. exponential random variables with unit mean. The channel gains do not change during each period but independently change along the periods. An recovery failure occurs if either data object is not properly stored or accumulated information at data collector is less than threshold $Q$.

Fig. \ref{fig:sim1} shows the exponential order of the recovery failure probability versus the number of storage nodes for various sum storage constraints. It is shown that the exponential order increases with the number of storage nodes and the increasing slope depends on the sum storage constraint. For example, when  $T=2$, the increasing slope of the exponential order is $\frac{1}{2}$ as noted in Remark \ref{remark:2}.
Contrary to the conventional multiple relay or antenna diversity order, the exponential order dose not increase with the number of storage nodes since  the gain provided by multiple storage nodes is hampered by limited storage capacity.  
In other words, adding storage node can improve reliability of storage operation but its effect is marginal without sufficient storage capacity.

Fig. \ref{fig:sim2} shows the effect of the sum storage constraint on the exponential order of the recovery failure probability when the number of storage nodes is fixed as $K=10$ or $K=200$.  As predicted in Corollary \ref{col:opt_d}, the exponential order increases and converges as the sum storage budget grows. According to Theorem \ref{thm:allo}, with the optimal storage allocation in high SNR, the storage size allocated to each node becomes smaller as the number of storage nodes increases. For example,
to obtain 80\% of the maximum exponential order which is marked as the black circles, the storage sizes required at each node (i.e., $T/K$) are  $a_i=1/40$ and $a_i=1/3$ for $K=200$ and $K=10$, respectively.
Fig. 4 reveals that if sum storage capacity is relatively small compared to the number of storage nodes,  only a small amount of increment in storage capacity can considerably increase the exponential order. This phenomenon is more noticeable when the number of storage node is large, which implies that full exponential order is achievable with only a very small storage budget per storage node if the number of storage nodes is large enough.

\begin{figure}[!t]
    \centerline{\psfig{figure=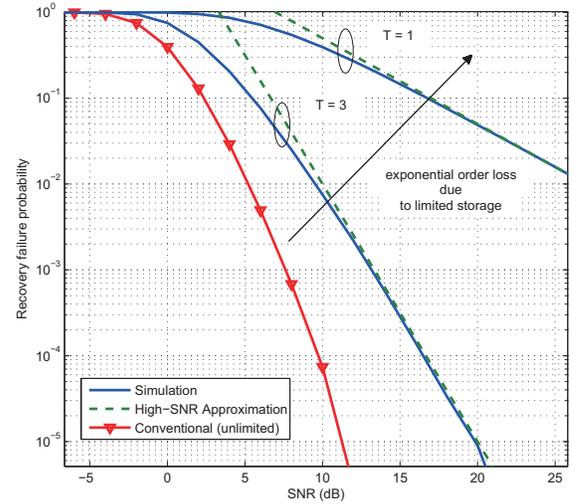,width=.85\columnwidth} }
    \caption{Recovery failure probabilities for $T=1, 3$, and $\infty$ when $K=5$ and their high SNR approximations}
    \label{fig:sim3}
\end{figure}

Fig. \ref{fig:sim3} figure verifies that the asymptotic analysis of recovery failure probability well approximate the recovery failure probability in high SNR regime. As shown in the analysis of exponential order, the recovery failure probability is degraded as the sum storage capacity is smaller. It is also verified that the high SNR approximation matches well with the simulation result if SNR is greater than 10 dB.
Given that a typical range of required SNR in 802.11n Wi-Fi is around from 10 to 30 dB
\cite{RX_sensitivity}, the high SNR approximation would be useful in performance evaluations .

\begin{figure}[!t]
    \centerline{\psfig{figure=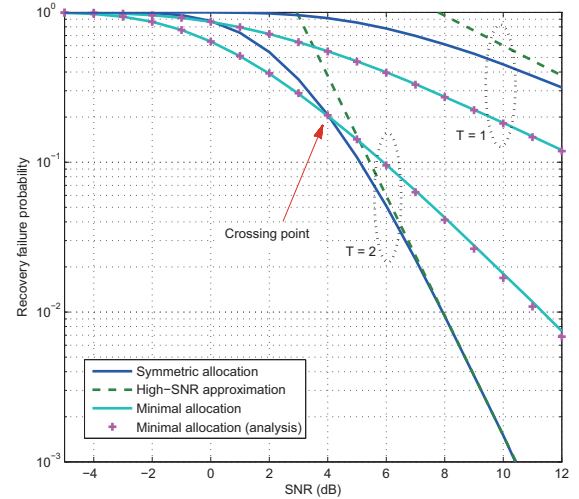,width=.85\columnwidth} }
    \caption{Recovery failure probabilities of the maximal symmetric allocation and the minimal allocation when $T=1,2$ and $K=6$. }
    \label{fig:sim4}
\end{figure}

\subsection{Low SNR Regime}
Fig. \ref{fig:sim4} exhibits that the minimal allocation is strictly better than the symmetric allocation in terms of the recovery failure probability in low SNR regime. For $T=1$,  although both of the allocation strategies have the same exponential order, the minimal allocation outperforms the symmetric allocation in terms of recovery failure probability. In the symmetric allocation, all the storage nodes have to decode the data object from the data source for the optimal performance, but this is unlikely in low SNR. Consequently, the performance of the symmetric allocation is restricted at the storing phase and worse than that of the minimal allocation in low SNR.  For $T=2$, the exponential orders of the symmetric allocation and the minimal allocation are $4$ and $2$, respectively. In high SNR, the symmetric allocation is definitely better but in low SNR, the recovery failure probability of the minimal allocation is much lower than that of the symmetric allocation. The crossing point between the two allocation schemes is around $4$ dB.

\subsection{Discussions on Storage Allocation in Intermediate SNR}

\begin{figure}[!t]
    \centerline{\psfig{figure=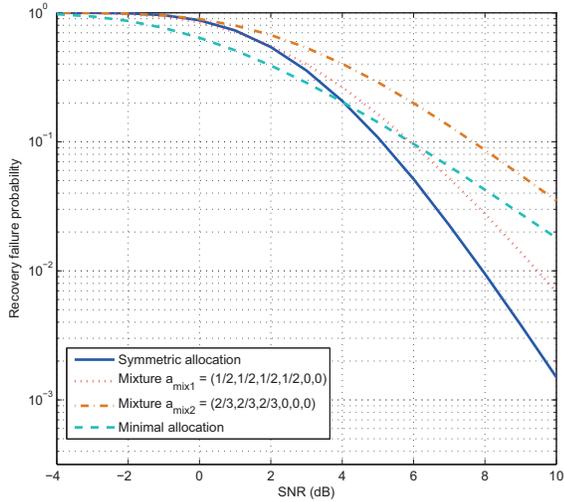,width=.85\columnwidth} }
    \caption{Recovery failure probability of various allocation strategies.}
    \label{fig:sim5}
\end{figure}

Although the optimal storage allocation is unknown for intermediate SNR regime, the optimal storage allocation strategies for high and low SNR regimes may suggest that a balanced allocation between the maximal symmetric allocation and the minimal allocation
be effective in intermediate SNR. For example, when there are 6 storage nodes and the sum storage budget is 2, i.e., $(K=6,\: T= 2)$, the symmetric allocation  is $\mathbf{a_{sym}}=\left(\frac{1}{3},\ldots,\frac{1}{3}\right)$ and the minimal allocation is $\mathbf{a_{min}}=\left(1, 1, 0, 0, 0, 0\right)$. Possible balanced allocation strategies between them could be $\mathbf{a_{mix1}}=\left(\frac{1}{2}, \frac{1}{2}, \frac{1}{2}, \frac{1}{2}, 0, 0\right)$ and  $\mathbf{a_{mix2}}=\left(\frac{2}{3}, \frac{2}{3}, \frac{2}{3}, 0, 0, 0\right)$. Both of the allocation strategies require at least 2 storage nodes to recover the data object but $\mathbf{a_{mix1}}$ can choose 2 among 4 storage nodes while $\mathbf{a_{mix2}}$ can choose 2 among 3 storage nodes. As a result, $\mathbf{a_{mix1}}$ is expected to outperform $\mathbf{a_{mix2}}$, which is verified in Fig. \ref{fig:sim5}. Comparing $\mathbf{a_{mix1}}$ with the maximal symmetric allocation and the minimal allocation, the minimal allocation is the best until SNR is around 4 dB at which the minimal allocation and the maximal symmetric allocation cross, whereas the maximal symmetric allocation is the best after the crossing. This result strongly implies that selection between the minimal allocation and the maximal symmetric allocation suffices.

\section{Conclusion}
In this paper, we introduced a new wireless distributed storage model with a sum storage capacity and investigated its performance in terms of the recovery failure probability. Using exponential order analysis, we proved that the maximal symmetric allocation is the optimal allocation strategy for high SNR regime. For the maximal symmetric allocation, we also presented an approximated representation of the recovery failure probability based on a high SNR approximation. On the other hand, using asymptotic analysis for low SNR,  the minimal allocation with $\lfloor T \rfloor$ complete storage nodes and one incomplete storage node was shown to be optimal in low SNR regime. If the sum storage capacity is given as an integer value, we derived the exact recovery failure probability of the minimal allocation in low SNR regime. Based on the numerical investigation, we also showed that a proper selection between the minimal allocation and the maximal symmetric allocation would make any balance allocation unnecessary.

\appendices
\def\thesection{\Alph{section}}%
\def\thesectiondis{\Alph{section}}%
\setcounter{equation}{0}
\renewcommand{\theequation}{A.\arabic{equation}}

\section{Proof of the Lemma \ref{lem:d_order}\label{app:1}}
\begin{figure}[t!]
         \begin{center}
         \includegraphics[scale=.7]{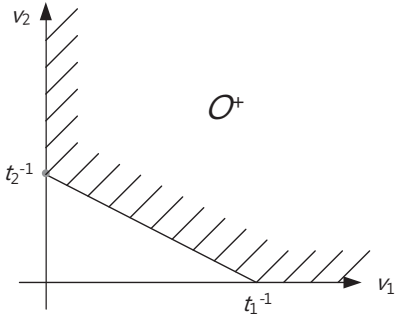}
         \caption{Exponential order when there are only two nodes. $O^+$ corresponds to  the dashed region. The exponential order is dominated and determined by $\min t_i^{-1}$. The exponential order is maximized when $t_1=t_2$ for the two node case. \label{fig:out_region}}
         \end{center}
        \end{figure}

By the law of total probability, the recovery failure probability is given by
  \begin{align}
    \text{Pr}_{f}[Q]=\sum_{\mathcal{D}\subseteq\{1,\ldots,K\}}\text{Pr}_{f}[Q\:|\:\mathcal{D}]\text{Pr}[\mathcal{D}]
  \end{align}
  where $Q$ is the rate of the data object, which is equivalent to the data object size. The probability for the decoding set $\mathcal{D}$ is
  obtained as
  \begin{align}
    \text{Pr}[\mathcal{D}]\nonumber
    &=\prod_{i\in\mathcal{D}}\text{Pr}\left[\log_2(1+\rho|h_{s,i}|^2)>Q\right]\nonumber\\
    &\times\prod_{i\in\{1,\ldots,K\}\setminus\mathcal{D}}\text{Pr}\left[\log_2(1+\rho|h_{s,i}|^2)<Q\right]\\
    &=\prod_{i\in\mathcal{D}}\text{Pr}\left[\rho|h_{s,i}|^2>2^Q-1\right]\nonumber\\
    &\times\prod_{i\in\{1,\ldots,K\}\setminus\mathcal{D}}\text{Pr}\left[\rho|h_{s,i}|^2<2^Q-1\right]\\
    &\overset{(a)}{\doteq}\prod_{i\in\mathcal{D}}\text{Pr}\left[\rho^{1-v_{s,i}}>\rho^0\right]\prod_{i\in\{1,\ldots,K\}\setminus\mathcal{D}}\text{Pr}\left[\rho^{1-v_{s,i}}<\rho^0\right]\\
     &\doteq\prod_{i\in\mathcal{D}}\rho^0\cdot
    \prod_{i\in\{1,\ldots,K\}\setminus\mathcal{D}}\rho^{-1}\\
    &=\rho^{-(K-|\mathcal{D}|)}\label{eq:second}
  \end{align}
  where $|\mathcal{D}|$ is the cardinality of the decoding set $\mathcal{D}$ and $v_{s,i}$ is the exponential order of $1/|h_{s,i}|$; $(a)$ comes from the definition of the exponential order. Assuming that $\{|g_i|^2\}$ are i.i.d. exponential random variables, for a given decoding set $\mathcal{D}$, the conditional recovery failure probability is upper bounded by
  \begin{align}
    \text{Pr}_{f}[Q\:|\:\mathcal{D}]=\: &\text{Pr}\left[\sum_{i\in\mathcal{D}}t_i\log_2(1+\rho|h_{i,c}|^2)<Q\right]
    \\&\mathbf{1}[\sum_{i\in\mathcal{D}}a_i\geq1]+\mathbf{1}[\sum_{i\in\mathcal{D}}a_i<1]\\
    &\overset{(a)}{\leq}\: \text{Pr}\left[\sum_{i\in\mathcal{D}}t_i\log_2(1+\rho|g_i|^2)<Q\right]\nonumber\\
    &\times\mathbf{1}[\sum_{i\in\mathcal{D}}a_i\geq1]+\mathbf{1}[\sum_{i\in\mathcal{D}}a_i<1]\\
    &\overset{(b)}{\doteq}\left\{
        \begin{array}{ll}
        \rho^{-\min_{i\in\mathcal{D}}{t_i}^{-1}}, &\textrm{for }\sum_{i\in\mathcal{D}}a_i\geq1,\\
        1, &\textrm{for } \sum_{i\in\mathcal{D}}a_i<1
        \end{array}
        \right.\label{eq:cond1}
    \end{align}
where $(a)$ is because  $|h_{i.c}|^2$ is replaced by $|g_i|^2$; the optimal transmit time allocation $\{t_i\}$ depends on the ordered channel gains of $|h_{i.c}|^2$, but not on
$|g_i|^2$, which yields higher recovery failure probability due to unoptimized transmit time allocation for $\{|g_i|^2\}$.  $(b)$ follows from the definition of the exponential order and
  \eqref{eq:expresult} (See \cite{Azarian:DDF} for more details) such that
  \begin{align}
\text{Pr}\left[\sum_{i\in\mathcal{D}}t_i\log(1+|g_i|^2\rho)<Q\right]&\doteq\rho^{-\inf_{\mathbf{v}\in O^+}\sum_{i\in\mathcal{D}}v_{i}}
\nonumber\\
&\doteq \rho^{-\min_{i\in\mathcal{D}}{t_i}^{-1}} \label{eq:example}
\end{align} where $O=\{v_{i} | v_{i}\in\mathcal{D} ,\sum_{i\in\mathcal{D}}t_i(1-v_{i})<0\}$ and $v_{i}$ is the exponential order of
$1/|g_{i}|^2$. Fig. \ref{fig:out_region} illustrate a simple example of \eqref{eq:example}  when there are only 2 nodes.

On the other hands, if the channel gains for each node are replaced by the highest channel gain among them during the recovery phase, a lower bound on the conditional recovery failure probability is obtained as
\begin{align}
    &\text{Pr}_{f}[Q\:|\:\mathcal{D}]\nonumber\\
    &\geq\:\text{Pr}[\log_2(1+\rho\max_{i\in\mathcal{D}}|h_{i,c}|^2)<Q]
    \mathbf{1}[\sum_{i\in\mathcal{D}}a_i\geq1]+\mathbf{1}[\sum_{i\in\mathcal{D}}a_i<1]\\
    &\doteq \rho^{-|\mathcal{D}|}\mathbf{1}\left[\sum_{i\in\mathcal{D}}a_i\geq1\right]
    +\mathbf{1}\left[\sum_{i\in\mathcal{D}}a_i<1\right]\label{eq:cond2}
     \end{align}
     
Combining
\eqref{eq:second}, \eqref{eq:cond1}, and \eqref{eq:cond2}, the exponential order of the recovery failure probability is lower and upper bounded as \eqref{thm:d_order1} and \eqref{thm:d_order2}, respectively.


\section{Proof of the Lemma \ref{lemma:3}\label{app:2}}

\begin{figure*}[h]
\begin{align}
    &\text{Pr}\left[(1-a_j^{\downarrow})\log\left(1+|h_{i,c}|^2\rho\right)+a_j^{\downarrow}\log\left(1+\max\{|h_{i,c}|^2,|h_{j,c}|^2\}\rho\right)>Q\right]\\
    &\overset{(a)}{\approx}\text{Pr}\left[(1-a_j^{\downarrow})|h_{i,c}|^2\rho+a_j^{\downarrow}\max\{|h_{i,c}|^2,|h_{j,c}|^2\}\rho>Q\right]\\
    &=1-\text{Pr}\left[(1-a_j^{\downarrow})|h_{i,c}|^2+a_j^{\downarrow}\max\{|h_{i,c}|^2,|h_{j,c}|^2\}\leq Q/\rho\right]\\
    &=1- \text{Pr}\left[(1-a_j^{\downarrow})|h_{i,c}|^2+a_j^{\downarrow} |h_{i,c}|^2 \leq Q/\rho ~\mid  |h_{i,c}|^2>|h_{j,c}|^2\right]\text{Pr}\left[|h_{i,c}|^2>|h_{j,c}|^2\right]\nonumber\\
    &\quad - \text{Pr}\left[(1-a_j^{\downarrow})|h_{i,c}|^2+a_j^{\downarrow} |h_{j,c}|^2 \leq Q/\rho ~\mid  |h_{i,c}|^2 \leq |h_{j,c}|^2\right]\text{Pr}\left[|h_{i,c}|^2 \leq |h_{j,c}|^2\right]\\
    &=1-\frac{1}{2} \left(1- \exp\left(-\frac{Q}{\rho}\right)\right) \nonumber\\
    &\qquad- \frac{1}{2}\int_0^{Q/\rho}\text{Pr}\left[(1-a_j^{\downarrow})t+a_j^{\downarrow}\max(t,|h_{j,c}|^2)<Q/\rho\right]\text{Pr}\left[|h_{i,c}|^2=t\right] \:dt\\
    &=1-\frac{1}{2} \left(1- \exp\left(-\frac{Q}{\rho}\right)\right)-\frac{1}{2}\int_0^{Q/\rho}\text{Pr}\left[|h_{j,c}|^2<\frac{Q/\rho-(1-a_j^{\downarrow})t}{a_j^{\downarrow}}\right]\exp(-t) \:dt\\
    &=1-\frac{1}{2} \left(1- \exp\left(-\frac{Q}{\rho}\right)\right)-\frac{1}{2}\int_0^{Q/\rho}\left(1-\exp\left(-\frac{Q/\rho-(1-a_j^{\downarrow})t}{a_j^{\downarrow}}\right)\right)\exp(-t) \:dt\\
    &=\exp\left(-\frac{Q}{\rho}\right)+\frac{a_j^{\downarrow}}{2(1-2a_j^{\downarrow})}\left(\exp\left(-2\cdot\frac{Q}{\rho}\right)-\exp\left(-\frac{Q}{\rho }\cdot \frac{1}{a_j^{\downarrow}}\right)\right)\label{eq:cvx}\\
    &=f_{\text{prob}}(a_j^{\downarrow})\label{eq:out_form}
\end{align}where $(a)$ comes from $\log(1+x)\approx x$ when $x$ is small. \\
\hrule
\end{figure*}

For the allocation $\mathbf{a}_{(K_1,K_2)}$ defined in the proof of Lemma \ref{lemma:2}, let  $K_1$ be $\lfloor T\rfloor$ and then, the sum of the storage sizes allocated to $K_2$ incomplete storage nodes is equal or less than $T-\lfloor T\rfloor$. There can be up to $K_2=K-K_1$ incomplete storage nodes in this storage allocation. Let this allocation be $\mathbf{a}_{(K_1,K_2)}'$ where $K_2=K-K_1$. Then the recovery probability  for $\mathbf{a}_{(K_1,K_2)}'$ is given by
\begin{align}
  &\bar{P}_O(\mathbf{a}_{(K_1,K_2)}')
    =\sum_{k_1=1}^{K_1}\binom{K_1}{k_1}\frac{e^{-k_1\cdot\frac{2^Q-1}{\rho}}}{\big(1-e^{-\frac{2^Q-1}{\rho}}\big)^{-(K_1+K_2-k_1)}}\label{eq:lemma3-11}\\
    &\times\text{Pr [ Recovery from $k_1$ out of $K_1$ com. nodes ]}\label{eq:lemma3-12}\\
    &+\sum_{k_1=1}^{K_1}\sum_{\mathcal{I}\neq \emptyset}\binom{K_1}{k_1}\frac{e^{-(k_1+|\mathcal{I}|)\cdot\frac{2^Q-1}{\rho}}}{\big(1-e^{-\frac{2^Q-1}{\rho}}\big)^{-(K_1+K_2-k_1-|\mathcal{I}|)}}\label{eq:lemma3-21}\\
    &\times\text{Pr [ Recovery from $k_1$ com. and inc. nodes in $\mathcal{I}$ ]}\label{eq:lemma3-22}
\end{align}
where $\mathcal{I}$ is a subset consists of the incomplete storage nodes in $\mathcal{D}$.
The first term consisting of \eqref{eq:lemma3-11} and \eqref{eq:lemma3-12} has a dominant scale in the recovery probability and is common regardless of $K_2$ in  $\mathbf{a}_{(K_1,K_2)}'$. Therefore, we have to focus on the second  term consisting of \eqref{eq:lemma3-21} and \eqref{eq:lemma3-22} to analyze the effect of incomplete storage allocations.

Now we prove that in the second term, the case when $k_1=1, |\mathcal{I}|=1$ leads to a dominant scale as $\rho\rightarrow 0$;
\begin{align}
    &\sum_{k_1=1}^{1}\sum_{|\mathcal{I}|=1}\binom{K_1}{k_1}e^{-2\cdot\frac{2^Q-1}{\rho}}\left(1-e^{-\frac{2^Q-1}{\rho}}\right)^{K_1+K_2-2}\nonumber\\
    &\times\text{Pr [ Recovery from $k_1(=1)$ com. and inc. nodes in $\mathcal{I}$ ]}\nonumber\\
    \overset{(a)}{\geq}&\sum_{|\mathcal{I}|=1}K_1e^{-2\cdot\frac{2^Q-1}{\rho}}\left(1-e^{-\frac{2^Q-1}{\rho}}\right)^{K_1+K_2-2}\nonumber\\
    &\times\text{Pr [ Recovery from 1 com. node ]}\nonumber\\
    \overset{(b)}{\gtrsim}&\: (K_1 K_2-\delta_1)e^{-2\cdot\frac{2^Q-1}{\rho}}e^{-\frac{2^Q-1}{\rho}}\quad\text{for arbitrary small } \delta_1\nonumber\\
    =&\Theta\left(\exp\left(-3\cdot\frac{2^Q-1}{\rho}\right)\right)\label{eq:l31}
\end{align}
where $(a)$ is from that recovery from 1 complete storage only without help of incomplete storage nodes yields worse recovery probability; $(b)$ is satisfied because $\big(1-e^{-\frac{2^Q-1}{\rho}}\big)^{K_1+K_2-2}$ goes to 1 as $\rho$ increases. For $|\mathcal{I}|\geq 2$, the followings hold:
\begin{align}
    &\sum_{k_1=1}^{K_1}\quad\sum_{|\mathcal{I}|\geq 2}\binom{K_1}{k_1}
    \frac{e^{-(k_1+|\mathcal{I}|)\cdot\frac{2^Q-1}{\rho}}}
    {\big(1-e^{-\frac{2^Q-1}{\rho}}\big)^{-(K_1+K_2-k_1-|\mathcal{I}|)}}\nonumber\\
    &\times\text{Pr [ Recovery from $k_1$ com. and inc. nodes in $\mathcal{I}$ ]}\nonumber\\
    \overset{(a)}{\leq}&\sum_{k_1=1}^{K_1}
    \frac{\sum_{|\mathcal{I}|\geq 2}\binom{K_1}{k_1}e^{-(k_1+|\mathcal{I}|)\cdot\frac{2^Q-1}{\rho}}}
    {\underbrace{\big(1-e^{-\frac{2^Q-1}{\rho}}\big)^{-(K_1+K_2-k_1-|\mathcal{I}|)}}_{\geq1}}\nonumber\\
    &\underbrace{\times\text{Pr [ Recovery from $k_1+|\mathcal{I}|$ com. nodes ]}}_{=1-(1-\exp(-(2^Q-1)/\rho))^{k_1+|\mathcal{I}|}\:\:\leq\:\: (k_1+|\mathcal{I}|)\exp(-(2^Q-1)/\rho)}\nonumber\\
    \overset{(b)}{\lesssim}&\sum_{k_1=1}^{K_1}\quad\sum_{|\mathcal{I}|\geq 2}\binom{K_1}{k_1}
    e^{-(k_1+|\mathcal{I}|)\cdot\frac{2^Q-1}{\rho}}
    \left(k_1+|\mathcal{I}|\right)e^{-\frac{2^Q-1}{\rho}}\nonumber\\
    =&\Theta\left(\exp\left(-4\cdot\frac{2^Q-1}{\rho}\right)\right)\label{eq:l32}
\end{align}
where $(a)$ is from that a recovery from $k_1+|\mathcal{I}|$ complete nodes is always better than that from $k_1$ complete storage nodes and $|\mathcal{I}|$ incomplete storage nodes; $(b)$ is because
\begin{align}
&\text{Pr [ Recovery from $k_1+|\mathcal{I}|$ com. nodes ]} \\
&=1-\left(1-\exp\left(-\frac{2^Q-1}{\rho}\right)\right)^{k_1+|\mathcal{I}|}\nonumber\\
&\leq \left(k_1+|\mathcal{I}|\right)\exp\left(-\frac{2^Q-1}{\rho}\right).\nonumber
\end{align} Comparing \eqref{eq:l32} with \eqref{eq:l31}, we verify that the case when $k_1=1$ and $ |\mathcal{I}|=1$ is dominant in the second term consisting of \eqref{eq:lemma3-21} and \eqref{eq:lemma3-22}.
Consequently, in low SNR regime, the optimal allocation strategy for the remaining $T-\lfloor T\rfloor$ budget  can be identified by solving following optimization problem:
\begin{align}
&\max_{a_j^{\downarrow}}\sum_{|\mathcal{D}\setminus\mathcal{I}|=1}\sum_{|\mathcal{I}|=1}\exp\left(-(|\mathcal{D}\setminus\mathcal{I}|+|\mathcal{I}|)\cdot\frac{2^Q-1}{\rho}\right)\nonumber\\
&\times\left(1\!-\!\exp\left(-\frac{2^Q-1}{\rho}\right)\right)^{K_1+K_2-|\mathcal{D}\setminus\mathcal{I}|-|\mathcal{I}|}\nonumber\\
&\times\text{Pr [ Recovery from 1 com. in $\mathcal{D}\setminus\mathcal{I}$ and 1 inc. nodes in $\mathcal{I}$ ]}\nonumber\\
&\text{subject to} \sum_{j=1}^{K_2}a_j^{\downarrow}=T-\lfloor T \rfloor \nonumber
\end{align}
where $a_j^{\downarrow}$ is the $j$th largest value among the storage sizes allocated to $K_2$ incomplete storage nodes, that is, the allocation to $K_2$ storage nodes is rewritten in descending order such that $\left(a_1^{\downarrow}\geq\cdots\geq a_{K_2}^{\downarrow}\right)$. Removing common terms and making the problem concise, we can reduce the optimization problem to
\begin{align}
  &\max_{a_j^{\downarrow}}\sum_{i=1}^{K_1}\sum_{j=1}^{K_2}\text{Pr [ Recovery from a com. $i$ and an inc. node $j$ ]}.\nonumber\\
  &=\max_{a_j^{\downarrow}}\sum_{i=1}^{K_1}\sum_{j=1}^{K_2}\text{Pr} \Big[(1-a_j^{\downarrow})\log\left(1+|h_{i,c}|^2\rho\right)\nonumber\\
  &\qquad\qquad +a_j^{\downarrow}\log\left(1+\max\{|h_{i,c}|^2,|h_{j,c}|^2\}\rho\right)>Q\Big]\\
  &\text{subject to} \sum_{j=1}^{K_2}a_j^{\downarrow}=T-\lfloor T \rfloor.
\end{align}where $|h_{i,c}|^2$ and $|h_{j,c}|^2$ are i.i.d. exponential random variables. Note that the probabilities for the summation are based on independent events and have the form in \eqref{eq:out_form}.
Note that \eqref{eq:cvx} is a convex function with respect to $a\in[0,1]$ since 1) it is an increasing function of $a\in[0,1]\setminus\{0.5\}$; 2) for $a=0.5$, the limit of \eqref{eq:cvx} exists and \eqref{eq:cvx} is continuous on $a\in[0,1]$; 3) for $0\leq a<0.5$, the dominant term is $\frac{a}{2(1-2a)}\exp\left(-2Q/\rho\right)$ and its first and second derivatives with respect to $a$ are always negative and positive, respectively, i.e., $\frac{-1}{2(1-2a)^2}<0$ and  $\frac{2}{(1-2a)^3}>0$; 4) for $0.5<a\leq 1$, the dominant term is $-\frac{a}{2(1-2a)}\exp\left(-\frac{Q}{\rho a}\right)$ which is also convex with respect to $a$ because we can adjust $\rho$ as small as we want in low SNR.

Consequently, $\sum_{i}\sum_{j}\text{Pr} \big[(1-a_j^{\downarrow})\log\left(1+|h_{i,c}|^2\rho\right)+a_j^{\downarrow}\log\left(1+\max\{|h_{i,c}|^2,|h_{j,c}|^2\}\rho\right)>Q\big]=K_1\sum_{j}f_{\text{prob}}(a_{j}^{\downarrow})$ is convex on a real interval $[0,1]$ because a sum of convex functions is also convex. Moreover, it is symmetric for $a_j^{\downarrow}$ and hence it is a Schur-convex function \cite{boyd:convex}. Therefore, with the constraint $\left(\sum_{j}a_j^{\downarrow}=T-\lfloor T \rfloor\right)$, the strong majorization holds for the allocation vector with only one non-zero element. That is, $a_1^{\downarrow}=T-\lfloor T\rfloor, a_2^{\downarrow}=\cdots a_{K_2}^{\downarrow}=0.$


\end{document}